\documentclass[sigconf, natbib=true]{acmart}

\AtBeginDocument{%
  }

\copyrightyear{2026}
\acmYear{2026}
\setcopyright{cc}
\setcctype{by}
\acmConference[CIKM '26]{Proceedings of the 35th ACM International Conference on Information and Knowledge Management}{November 07--11, 2026}{Rome, Italy}
\acmBooktitle{Proceedings of the 35th ACM International Conference on Information and Knowledge Management (CIKM '26), November 07--11, 2026, Rome, Italy}
\acmDOI{10.1145/3799682.3839253}
\acmISBN{979-8-4007-2539-5/2026/11}

\begin{document}

\title[Phased Agentic Workflow]{A Phased Workflow for Operating LLM-Based Coding Agents}

\author{Ante Kapetanovic}
\orcid{0000-0001-5507-5788}
\email{ankapetanovic@infobip.com}
\affiliation{%
  \institution{Infobip}
  \city{Split}
  \country{Croatia}
}

\author{Tomislav Duricic}
\orcid{0000-0002-7229-9596}
\email{tduricic@infobip.com}
\affiliation{%
  \institution{Infobip}
  \city{Zagreb}
  \country{Croatia}
}

\author{Andro Mercep}
\orcid{0000-0003-1229-3149}
\email{amercep@infobip.com}
\affiliation{%
  \institution{Infobip}
  \city{Zagreb}
  \country{Croatia}
}

\author{Emanuel Lacic}
\orcid{0000-0002-3059-0502}
\email{emlacic@infobip.com}
\affiliation{%
  \institution{Infobip}
  \city{Zagreb}
  \country{Croatia}
}

\renewcommand{\shortauthors}{Ante Kapetanovic, Tomislav Duricic, Andro Mercep, and Emanuel Lacic}



\begin{abstract}
  LLM-based coding agents combine a foundation model with a harness that shapes agent behavior. For non-trivial tasks, how practitioners structure their work with the coding agents determines whether reliable results follow. We report on a phased workflow for operating coding agents developed by the AI research team at Infobip. The workflow structures agent-assisted development into four phases where human effort is front-loaded and delegation increases as artifacts mature. Context management is the central concern, addressed through four strategies applied at each phase to counter known failure modes. From practitioner experience, we observe that upstream errors in research and planning can compound across later phases, while correcting generated code can introduce bloat and fragility. This motivates front-loading human review. We identify two open problems: the absence of metrics for workflow effectiveness and the gap between formalized context management components and the workflow-level patterns that practitioners need.
\end{abstract}


\maketitle

\section{Introduction}

LLM-based agents have been broadly applied across software engineering tasks~\cite{liu2026llmagentsse}, but their effectiveness on non-trivial tasks depends less on the choice of foundation model than on how practitioners configure the agent's harness.
B\"{o}ckeler~\cite{bockeler2026harness} distinguishes the \emph{provider harness}, built into the agent by its vendor, from the \emph{user harness}, which practitioners configure through project-level instructions~\cite{gloaguen2026agentsmd}, tool selection, session structure, and context management.
Performance degrades consistently with input length even on trivial tasks~\cite{hong2025contextrot}, multi-turn settings show an average 39\% performance drop~\cite{laban2025lostturn}, and irrelevant context reduces accuracy even when the correct answer is present~\cite{hong2025contextrot}. Overall, unstructured use of AI coding tools can reduce developer productivity~\cite{becker2025metr}.
These findings motivate structured approaches to agentic development.

We report on a phased workflow for operating coding agents at Infobip,\footnote{Workflow materials: \url{https://github.com/infobip/agentic-development-workflow}} a cloud communications platform that is supported by around 900 engineers and processes over 42 billion messages monthly across more than 190 countries.
Extending the research-plan-implement paradigm~\cite{horthy2025ace}, the workflow structures \sloppy{agent-assisted} development into four phases where human effort decreases as artifacts mature.
In our experience, flawed research can propagate into flawed plans and code, while correcting generated code can introduce bloat and fragility.
This motivates front-loading human review.
Our contributions are thus: (1)~a phased workflow for operating coding agents, (2)~concrete practices for configuring the user harness at each phase, and (3)~introduction of two open problems for the research community.

\section{Phased Workflow}

Our proposed workflow consists of four phases: research, planning, task definition, and implementation.
Research and planning are performed once per project; task definition and implementation repeat as tasks are executed and refined.
Human involvement is highest in the upstream phases, where errors carry the greatest cost, and decreases as artifacts mature.
When implementation reveals poorly defined tasks, the practitioner returns to task definition instead of patching the implementation.

At each phase, the practitioner manages context through four strategies described by Martin~\cite{martin2025contexteng}: \emph{write} (persist information outside the context window), \emph{select} (retrieve relevant information into the window), \emph{compress} (retain only required tokens), and \emph{isolate} (split context across sessions or agents).
These strategies counter failure modes identified by Breunig~\cite{breunig2025longcontexts}: \emph{distraction} (over-reliance on accumulated patterns), \emph{confusion} (irrelevant content competing for attention), \emph{poisoning} (hallucinations persisting across turns), and \emph{clash} (new information conflicting with existing context).
The relationship is not one-to-one: isolating context in fresh sessions prevents both poisoning and distraction, while compressing findings into a summary reduces both confusion and distraction.
Each phase applies strategies matched to its specific risks.

\paragraph{Research.} The practitioner and agent explore the problem space together.
The agent operates with access limited to read operations, codebase search, and domain-specific tools such as literature search.
The practitioner guides research direction but withholds opinions to avoid anchoring the agent's exploration.
All findings are validated before proceeding.
This phase selects context by loading only research-relevant tools and isolates the session to a focused scope, thus reducing the risk of confusion from irrelevant tool definitions and distraction from unrelated context.
The output is a curated research document that captures validated findings.

\paragraph{Planning.} Using the research document as input, the practitioner and agent develop an implementation approach through iterative discussion.
The practitioner works in the agent's default operating mode rather than a dedicated planning mode.
We found that dedicated planning modes push the agent to finalize prematurely, producing plan artifacts before the problem is sufficiently understood.
When the approach is settled, the practitioner reduces the discussion to validated decisions and writes them to a persistent artifact that replaces the full conversation in subsequent phases.
The act of manual curation is the primary application of the compress strategy in the workflow: it discards abandoned lines of reasoning and retains only what subsequent phases need, preventing poisoning from earlier missteps.

\paragraph{Task definition.} Each machine-readable task records a stable identifier, goal, dependencies, relevant code paths, acceptance criteria, and validation method.
Tasks are sized to complete in a fresh session and split when the required context would exceed the agent's effective operating range, which can be substantially below nominal capacity~\cite{kuratov2024babilong,hong2025contextrot}.

\paragraph{Implementation.} Each task runs in a fresh session loaded with project-level instructions and the task definition.
The task registry stores status, dependencies, acceptance criteria, and the latest outcome; an append-only activity log records actions, validation results, changed files or commits, failures, and the next action.
Git stores code state, while the registry and log store workflow state, making conversation state disposable.
After an interruption, a fresh session reads these artifacts, inspects the working tree and latest commit, and resumes or reverts incomplete work.
The agent then implements, validates, logs, updates task status, commits, and terminates under manual supervision or in an autonomous loop.

Project-level instructions are hand-crafted and evolve over time.
When the practitioner observes the agent repeating a mistake across sessions, a targeted correction is added.
Since instruction-following accuracy degrades with instruction count~\cite{jaroslawicz2025instructions}, corrections are kept to concise single-line entries.
The practitioner reviews output rather than writing it, but retains ownership of the overall process.

\section{Open Problems}

Two problems remain unresolved.
First, there are no established metrics for workflow effectiveness.
Practitioners cannot measure whether their phased workflow improves outcomes compared to unstructured delegation, whether task granularity is appropriate, or whether instructions are well-calibrated.
Existing coding agent benchmarks have known limitations that make them unreliable proxies for workflow effectiveness.
Candidate metrics might include task success rate per session, instruction correction rate, and context utilization at task completion, but these remain unexplored.

Second, research formalizes individual components of context management, such as pruning, compression, and tool selection, but not the workflow-level patterns that compose them.
Zhang et al.~\cite{zhang2025ace} demonstrate that structured context management yields measurable gains on agent benchmarks, but these gains are measured at the component level, not across practitioner workflows.
The absence of shared terminology makes it difficult to compare or transfer approaches across teams and domains.

\begin{acks}
This work was supported in part by the Infobip Global Communication Platform project (PK.1.1.07.0001), conducted within the Important Project of Common European Interest on Next Generation Cloud Infrastructure and Services (IPCEI-CIS).
\end{acks}

\section*{Speaker Bio}
Ante Kapetanovic is a Senior Researcher at Infobip, where he works on human-AI interaction, conversational AI and information retrieval, and trustworthiness.

\section*{Relevance}
This proposal addresses three CIKM Industry Day topics: system design from industry practitioners (a phased workflow for operating coding agents), domain-specific challenges (context management in agent-assisted development), and collaborative investigation between industry and academia (open problems on workflow metrics and practitioner patterns).

\section*{GenAI Usage Disclosure}
Generative AI tools were used to support the conceptualization of this article and to assist in drafting the accompanying repository. Google Scholar Labs was used to discover relevant references.

\bibliographystyle{ACM-Reference-Format}
\bibliography{references}

\end{document}